\documentclass[runningheads]{llncs}
\usepackage[T1]{fontenc}

\usepackage{booktabs}
\usepackage{multirow}
\usepackage{adjustbox}
\usepackage{subcaption}
\usepackage{todonotes}

\usepackage{amsmath}

\DeclareMathAlphabet{\mathmybb}{U}{bbold}{m}{n}

\usepackage[pagebackref=true,breaklinks=true,colorlinks,bookmarks=false]{hyperref}

\usepackage{graphicx}
\usepackage[misc]{ifsym} 
\begin{document}
\title{Detecting respiratory motion artefacts for cardiovascular MRIs to ensure high-quality segmentation}
%
\titlerunning{Detecting motion artefacts for ensuring HQ CMR segmentation}

\author{Amin Ranem\inst{1}$^*$ \orcidID{0000-0003-0783-6903}\textsuperscript{(\Letter)} \and
John Kalkhof\inst{1}$^*$ \and
Caner Özer\inst{2}$^*$
\and
Anirban Mukhopadhyay\inst{1}
\and
Ilkay Oksuz\inst{2}}

\authorrunning{A. Ranem et al.}

\institute{Darmstadt University of Technology, Karolinenpl. 5, 64289 Darmstadt, Germany \email{amin.ranem@gris.informatik.tu-darmstadt.de} \and
Computer Engineering, Istanbul Technical University}

\def\thefootnote{*}\footnotetext{These authors contributed equally to this work}\def\thefootnote{\arabic{footnote}}

%
%

\maketitle              
\begin{abstract}
While machine learning approaches perform well on their training domain, they generally tend to fail in a real-world application. In cardiovascular magnetic resonance imaging (CMR), respiratory motion represents a major challenge in terms of acquisition quality and therefore subsequent analysis and final diagnosis. We present a workflow which predicts a severity score for respiratory motion in CMR for the CMRxMotion challenge 2022. This is an important tool for technicians to immediately provide feedback on the CMR quality during acquisition, as poor-quality images can directly be re-acquired while the patient is still available in the vicinity. Thus, our method ensures that the acquired CMR holds up to a specific quality standard before it is used for further diagnosis. Therefore, it enables an efficient base for proper diagnosis without having time and cost-intensive re-acquisitions in cases of severe motion artefacts. Combined with our segmentation model, this can help cardiologists and technicians in their daily routine by providing a complete pipeline to guarantee proper quality assessment and genuine segmentations for cardiovascular scans. The code base is available at \url{https://github.com/MECLabTUDA/QA_med_data/tree/dev_QA_CMRxMotion}.

\keywords{Cardiovascular MRI \and Respiratory motion artefact detection \and Semantic segmentation \and Image Quality Assessment.}
\end{abstract}
\section{Introduction}
\begin{figure}[]
\centering
\begin{minipage}[b]{0.87\linewidth}
\centerline{\includegraphics[width=\linewidth]{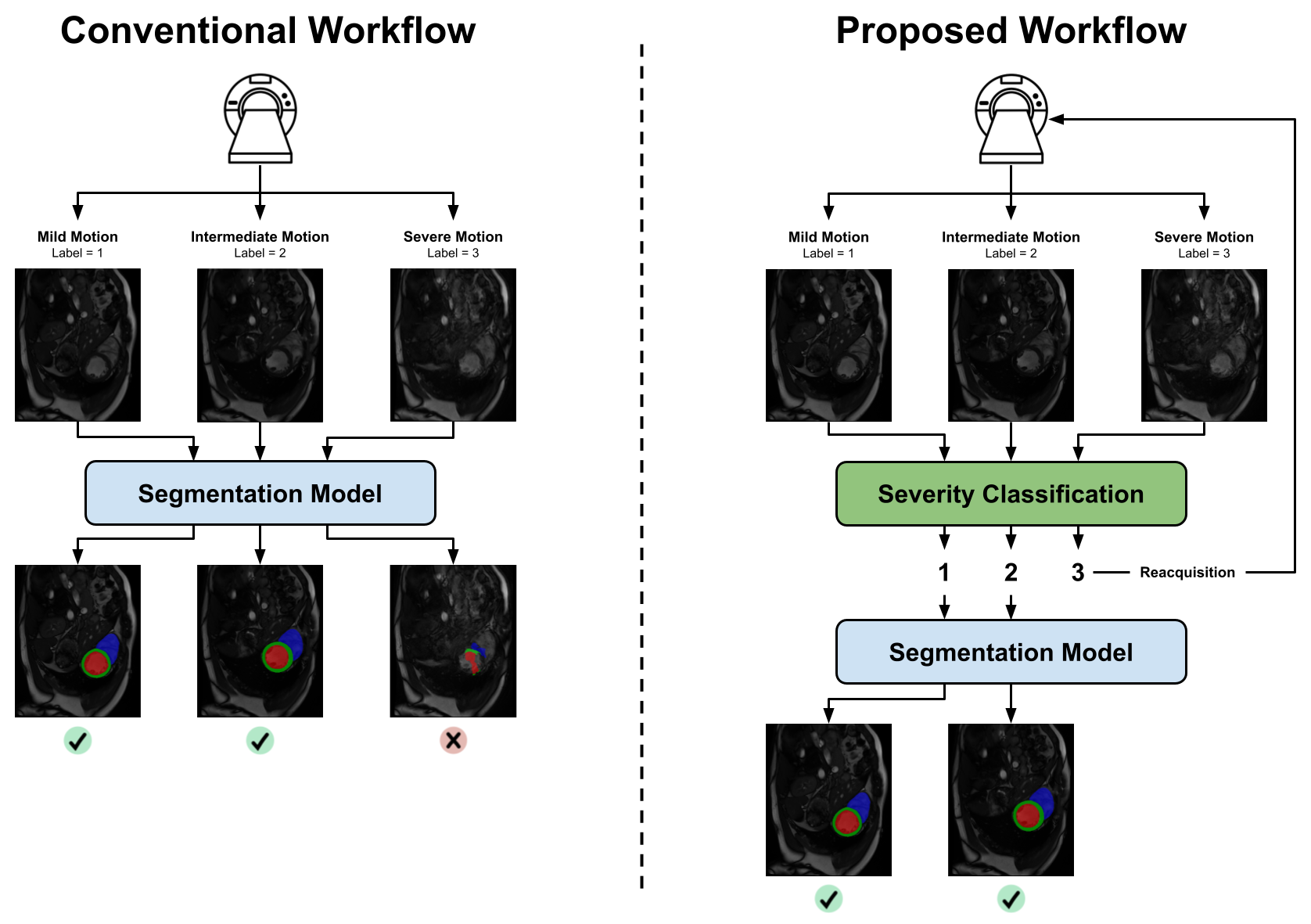}}
\end{minipage}
\caption{Segmentation models can generally deal with different levels of motion artefacts, although they break down at a certain point. Therefore, detecting when this breakdown occurs due to severe motion artefacts is essential to prevent the model from making incorrect predictions.
}
\label{fig:problem_definition}
\end{figure}

Respiratory motion artefacts are a common problem when performing image segmentation on cardiovascular magnetic resonance images (CMR). These motion artefacts can be more or less severe depending on the patient's ability to hold their breath. Over the past decades, there have been plenty of approaches trying to detect \cite{Ferreira2013,Karakamis2021,Oksuz2018,Oksuz2019} and reduce \cite{King_2012,Ozer2021a,Seppenwoolde_2002,Sinclair_2017,Zhang_2010,Oksuz2019a,Oksuz2020a} the artefacts produced by the patient's breathing patterns. These procedures are necessary since the breathing artefacts make anatomical boundaries unclear \cite{Zhang_2010}, and therefore, segmentation of them is more challenging or even impossible (see Fig. \ref{fig:problem_definition}). Furthermore, these high-quality segmentations play a critical role, especially in treatments like image-guided radiation \cite{Zhang_2010}, as otherwise healthy tissue could be damaged \cite{Seppenwoolde_2002}. 

Recent advances in deep learning help to perform high-quality cardiovascular segmentation on MRI images that contain mild to intermediate respiratory artefacts. However, it becomes much more difficult when the severity turns too strong. To ensure that the model always achieves sufficient performance, a classification model can be used as a first step in whether a cardiovascular MRI image has been acquired sufficiently well to perform segmentation.

\vspace{-0.3cm}
We propose combining a segmentation model and a respiratory motion classifier as part of the 'Extreme Cardiac MRI Analysis challenge under Respiratory Motion (CMRxMotion)' registered in MICCAI 2022. The first step is to use the classifier to predict whether the given MRI image has sufficiently mild respiratory artefacts for our segmentation model to produce high-quality segmentations. Then, after filtering out low-quality scans, segmentation is performed on the remaining samples.

Our proposed classification model achieves $67.5\%$ accuracy in classifying the severity of cardiovascular MRI images. The segmentation model then reaches an average Dice accuracy of $86.18\%$ on the remaining images containing mild to intermediate motion artefacts.
Our contributions are two-fold and can be summarised as follows:
\begin{itemize}
    \item[$\bullet$] We introduce an ordinal regression-based pipeline that reaches $67.5\%$ accuracy in predicting the severity of respiratory motion in MRI images. 
    \item[$\bullet$] We produce a segmentation model achieving $86.18\%$ Dice accuracy robust to images with mild to intermediate respiratory motion artefacts.
\end{itemize}





\section{Method}
We describe the components of our proposed workflow defined in Fig. \ref{fig:problem_definition}. First, the respiratory motion classifier (in Sec. \ref{sec:classification}) predicts the severity of CMRs. After the images with severe motion artefacts are omitted, the remainder is segmented with our segmentation model (in Sec. \ref{sec:segmentation}).

\subsection{Image Quality Assessment of Respiratory Motion Artefacts}
\label{sec:classification}
We propose a Cardiac MRI motion artefact identification system to distinguish the scans with mild, intermediate and severe motion artefacts. Although it is possible to use a generic classification framework with this objective, which uses a Softmax classifier at the end, we would also like to involve the relative label information between the artefact levels so that we can apply more supervision while training our models. In this regard, we adopt the work of Cao et al. (CORAL) \cite{Cao2020} and Shi et al. (CORN) \cite{Shi2021} to predict the artefact level of a medical scan while considering the rank consistency among predictions. 


\subsubsection{Rank Consistent Neural Networks:}
\begin{figure}
    \centering
    \includegraphics[scale=0.40]{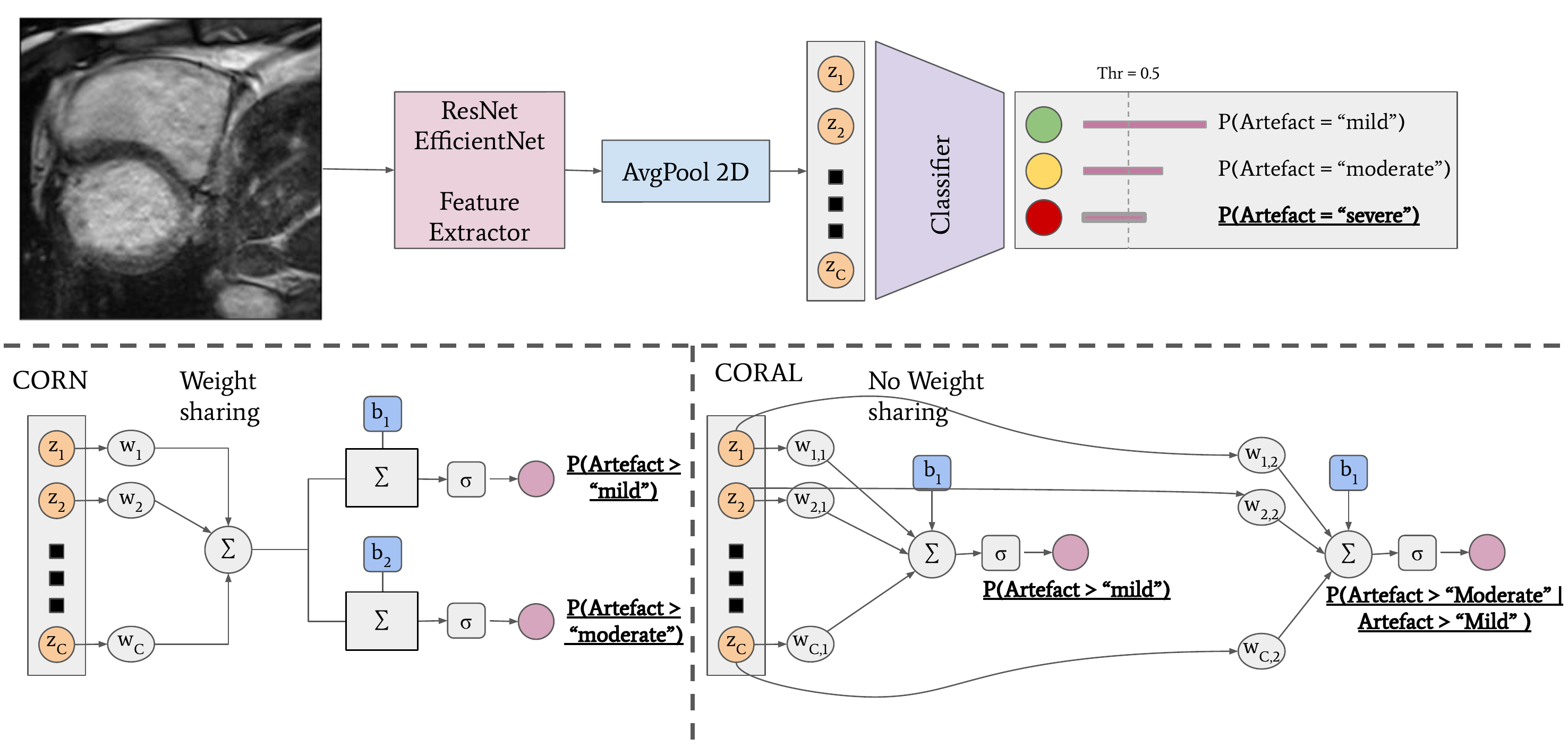}
    \caption{CORAL and CORN approaches for satisfying rank consistency.}
    \label{fig:rank_consistency}
\end{figure}

For both CORAL and CORN, suppose we have a deep neural network composed of a feature extractor and a classifier designed to process 2D image slices. In addition, let us suppose that we have a dataset $\mathcal{D}=\{x^{(i)}, y^{(i)}\}$ where $x^{(i)}$ is the $i^{th}$ training sample, and $y^{(i)}$ corresponds to its label.
Ordinal regression aims to minimise the cost function $L(r)$, where the mapping $r: \mathcal{X} \xrightarrow{} \mathcal{Y}$ is called a ranking rule, such that each label has a level. The overall framework is shown in Fig. \ref{fig:rank_consistency}. 

\subsubsection{CORAL:}

In order to train a CORAL model, we start with extending the class label of a sample. For instance, if there are $3$ classes in total, the binarised vector representation will be $[0, 0]$ for class 1, $[1, 0]$ for class 2, and $[1, 1]$ for class 3, different than the one-hot encoding scheme. Supposing that there exist $K$ ordinal classes, we can model them by using $K-1$ binary classifiers where we can predict the rank index of the sample as follows:

\begin{equation}
    q = 1 + \sum_{k=1}^{K-1} f_k(\textbf{z}_i).
    \label{eq:rank_index}
\end{equation}

In Eq. \eqref{eq:rank_index}, $f_k(.)$ is the thresholded output of the $k^{th}$ binary classifier, and $\textbf{z}_i$ is the vector output of the feature extractor of the $i^{th}$ sample, or in other words, $\textbf{z}_i = g(x_i; \textbf{W})$ where $\textbf{W}$ denotes the parameters of the feature extractor. 
This equation has a rank monotonicity assumption that, in order to predict the rank index of the image, all of the binary classifier outputs until some level should output $1$ whereas after that particular level, the output should be $0$. 
What is special about CORAL is that, the rank consistency is maintained by sharing the weight parameters of a classifier which are denoted by \textbf{w}.
However, each of the binary classifiers has different bias units, and each of them are denoted as $b_k$ (\textbf{b} in the vector form) to differentiate each of the predicted labels. 
As a consequence, we can define the probability value of a classifier $k$ as $\hat{P}(y^{(i)}_k = 1) = \sigma(\textbf{wz}_i + b_k)$ where $\sigma$ is the Sigmoid activation function.


This network is trained by minimising the weighted cross-entropy loss function in Eq. \eqref{eq:coral_loss} where $\lambda^{(k)}$ denotes the importance of each of the binary classifiers. 



\begin{equation}
        L(\textbf{w}, \textbf{W}, \textbf{b}) = - \sum_{i=1}^N \sum_{k=1}^{K-1} \lambda^{(k)} [log(\hat{P}(y^{(i)}_k = 1)) y_k^{(i)} + log(1 - \hat{P}(y^{(i)}_k = 1)) (1 - y_k^{(i)})]
    \label{eq:coral_loss}
\end{equation}

After training the network and obtaining the probability values, the final decision of the classifier $k$ is obtained by 

\begin{equation}
    f_k(\textbf{z}_i) = \mathmybb{1}(\hat{P}(y^{(i)}_k = 1) > 0.5),
    \label{eq:coral_predict}
\end{equation}

where $\mathmybb{1}(\cdot)$ is the indicator function. The overall decision for a sample is called rank index, which is defined in Eq. \eqref{eq:rank_index}.

\subsubsection{CORN:}
In contrast to CORAL, CORN argues that each of the binary classifier outputs is conditioned on the previous one. Supposing that the binary classifier's output is expressed as $f_k(\textbf{z}_i) = \hat{P}(y^{(i)} > y_k | y^{(i)} > y_{k-1} )$, the marginal probability of output random variable being greater than the level of $y_k$ is 

\begin{equation}
    \hat{P}(y^{(i)} > y_k) = \prod_{j=1}^{k} f_j(\textbf{z}_i).
    \label{eq:corn_conditional}
\end{equation}



Finally, to train CORN, we generate conditional training subsets, $S_k$, and optimise Eq. \eqref{eq:corn_loss}. Prediction is performed as in Eq. \eqref{eq:coral_predict}.

\begin{multline}
    L(\textbf{w}, \textbf{W}, \textbf{b}) = - \frac{1}{\sum_{j=1}^{K-1} |S_j|} \sum_{j=1}^{K-1} \sum_{i=1}^{|S_j|} [log(f_j(x^{(i)})) \mathmybb{1}(y^{(i)} > y_k) + \\ log(1 - f_j(x^{(i)})) \mathmybb{1}(y^{(i)} \leq y_k)]
    \label{eq:corn_loss}
\end{multline}


%



\subsection{CMR Image Segmentation with Realistic Respiratory Motion}
\label{sec:segmentation}
Creating high-quality segmentation for three-label CMR scans requires the network to be robust against respiratory motion artefacts. We tackle the segmentation problem of the left (LV) and right ventricle (RV) blood pools along with the left ventricular myocardium (MYO) with the nnU-Net framework \cite{isensee2018nnu}. nnU-Net is used in a two-dimensional (2D) and three-dimensional (3D) setup. Its dynamic framework design performs all relevant pre- and postprocessing steps. Thus, the CMRxMotion dataset is not resampled, cropped or normalised for the segmentation task. Based on the dataset specifics, the adaptive framework configures a corresponding U-Net making the framework state-of-the-art for several medical segmentation challenges \cite{isensee2019nnu}. The recently published ViT U-Net V2 (2D) \cite{ranem2022continual} from the Lifelong nnU-Net Framework \cite{gonzalez2022lifelong} is a nnU-Net based architecture with a base Vision Transformer (ViT) \cite{dosovitskiy2020image} backbone and achieves state-of-the-art segmentation performances for medical segmentation while introducing the successful mechanism of self-attention. The ViT U-Net network is used with the assumption that the attention mechanism of transformers can be leveraged to concentrate the self-attention on the motion artefact/heart area leading to a more robust segmentation network in severe artefact cases.

Three different architectures were analysed to create a robust segmentation network for CMR scans. We utilise the same experimental setup for all three architectures. The networks are trained on a random split for 250 epochs while using the pre-processing steps of the nnU-Net framework. The networks are trained on a system with 256GB DDR4 SDRAM, 2 Intel Xeon Silver 4210 CPUs and 8 NVIDIA Tesla T4 (16 GB) GPUs.


\section{Results}
We showcase the experimental results of the classifier (Sec. \ref{eval:classifier}) and the segmentation (Sec. \ref{eval:segmentation}) from  Synapse platform on the validation set in this section.

\subsection{Image Quality Assessment of Respiratory Motion Artefacts}
\label{eval:classifier}
We evaluate the performance using classification accuracy and Cohen's Kappa as suggested by CMRxMotion challenge. In Table \ref{tab:cls_res}, the official validation results for CMRxMotion challenge for $4$ trained models are provided. The first model utilises a ResNet-152 backbone, trained with SoftMax loss, while the rest of the models utilise an EfficientNet-B5 backbone \cite{Tan2019} which are trained by using focal loss \cite{Lin2020}, CORAL \cite{Cao2020} and CORN \cite{Shi2021}. We use Optuna \cite{Akiba2019} to tune the learning rate and find the best-performing optimiser by holding $4$ of the patients as the unofficial validation data. We re-ran the training 50 times and pruned it whenever the performance seemed insufficient over the course of iterations. 

\begin{table}[htb!]
\centering
\begin{tabular}{lccccc}
\toprule
\begin{tabular}[c]{@{}l@{}}Model\\ Name\end{tabular} & \begin{tabular}[c]{@{}c@{}}Focal\\ Loss\end{tabular} & CORAL & CORN & Acc $\uparrow{ }$  & \begin{tabular}[c]{@{}c@{}}Cohen's\\ Kappa $\uparrow{ }$\end{tabular} \\ \hline \hline
ResNet-152 & - & - & - & 0.55 & 0.307 \\
\hline
\multirow{3}{*}{EfficientNet-B5}                     & +                                                    & -     & -    & 0.650  & 0.416                                                   \\ 
                                                     & -                                                    & +     & -    & \textbf{0.675} & \textbf{0.451}                                                   \\ 
                                                     & -                                                    & -     & +    & 0.650      &  0.421                                                       \\ 
\bottomrule
\end{tabular}
\vspace{0.3 cm}
\caption{Performance of the three EfficientNet-B5 versions trained on the CMRxMotion data, evaluated on the official validation dataset. Bold values indicate the best performance among the three networks.}
\label{tab:cls_res}
\end{table}

\vspace{-0.75 cm}
Considering the numerical results, we see a positive impact of using the EfficientNet-B5 over the ResNet-152 model. Furthermore, if we compare the Efficient-Net-B5 models, we see a slight difference between the methods even though the figures are in favour of the CORAL model. 

\subsection{CMR Image Segmentation with Realistic Respiratory Motion}
\label{eval:segmentation}
The analysis of the segmentation networks is split into two consecutive parts; quantitative and qualitative analysis. All results are extracted based on the provided CMRxMotion validation dataset and the Synapse leaderboard, where ground truth labels are unknown to the participants.

\subsubsection{Quantitative Analysis:}
We evaluate the robustness of the segmentation models with the S\o{}rensen–Dice (Dice) and Hausdorff (HD95) metrics. Table \ref{tab:seg_res} summarises the results for the three trained networks using the validation dataset from CMRxMotion\footnote{The scores are extracted from the Synapse leaderboard of the challenge.}.

\begin{table}[htb!]
\centering
\begin{adjustbox}{width=0.86\linewidth}
{\begin{tabular}{lcccccccc}
\toprule 
\multirow{2}{*}{Architecture} && \multicolumn{3}{c}{Dice $\uparrow{ }$} && \multicolumn{3}{c}{HD95 $\downarrow{ }$} \\ 
 && LV & MYO & RV && LV & MYO & RV \\ \midrule \midrule
 nnU-Net$_{\text{2D}}$ && 0.8927 & 0.8038 & 0.8427 && 9.9391 & 5.3801 & 7.2047 \\
 \textbf{ViT U-Net}$_{\text{2D}}$ && \textbf{0.9003} & \textbf{0.8134} & \textbf{0.8718} && \textbf{9.8360} & \textbf{4.8026} & \textbf{6.4533} \\ \hline
 nnU-Net$_{\text{3D}}$ && 0.8589 & 0.7811 & 0.8545 && 12.549 & 6.2128 & 8.6875 \\
\bottomrule
\end{tabular}}
\end{adjustbox}
\vspace{0.3 cm}
\caption{Performance of the three nnU-Net versions trained
on the CMRxMotion data, evaluated on the corresponding validation dataset. Bold values indicate the best performance among the three networks.}
\label{tab:seg_res}
\end{table}

Analysing the table, one can deduce that the ViT U-Net achieves the best scores for every segmentation label leading to a more robust version than the plain nnU-Net. These results confirm the assumption that transformers' self-attention can be leveraged to create a more robust network for CMR segmentation tasks.

\subsubsection{Qualitative Analysis:}
\begin{figure}[h!]
\begin{center}
\includegraphics[width=.75\linewidth]{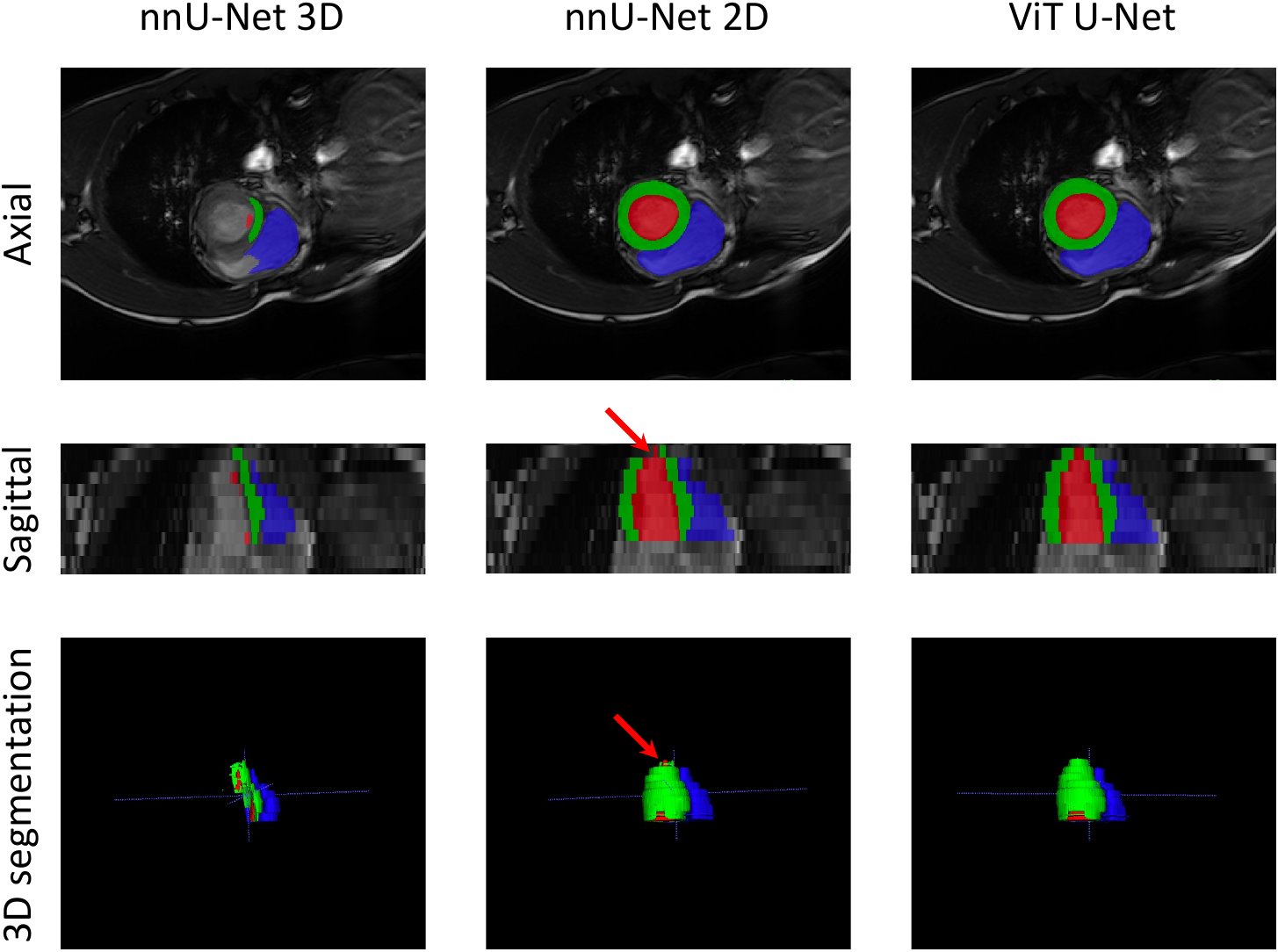}
\caption{Comparison of predictions between 3D/2D nnU-Net and ViT U-Net using a random CMRxMotion image from the validation set. The first row shows a random axial slice, the second row is a random sagittal slice and the last row is the 3D segmentation mask. Red arrows indicate the incomplete segmentations for nnuU-net, where ViT U-Net performs better.}
\label{fig:preds_seg}
\end{center}
\end{figure}

To further analyse the segmentation performance of each model, a comparison between the predictions for a random scan is illustrated in Figure \ref{fig:preds_seg}. As the random scan is from the validation set, we cannot provide a ground truth mask. 


\vspace{-0.75cm}
Closely analysing Figure \ref{fig:preds_seg}, one can easily see that there are differences in terms of robust segmentation masks among the three different architectures. The axial view clearly shows the lack of the RV (red) segmentation from the 3D nnU-Net, whereas no discernible difference can be seen between the 2D nnU-Net and ViT U-Net. The 3D segmentation masks show a significant difference between 3D nnU-Net and 2D nnU-Net/ViT U-Net regarding segmentation and robustness. When focusing on the top area of the sagittal view images (see red arrows), it is visible that the ViT U-Net segmentation for MYO and LV (green and red) is more complete than the one from the 2D nnU-Net, which is not that easy to spot in the 3D segmentation masks. This confirms the assumption made during the quantitative analysis that transformers lead to a more robust network.

\section{Discussion}
Detecting respiratory motion artefacts in cardiovascular images is not an easy task. Especially the inaccurately defined severity groups make it difficult to train a model that learns the different levels of severity. Our experiments have shown that well-known classification models like ResNet-152 \cite{he2016deep} do not perform very well in predicting the severity of these images. Therefore, we had to implement a more sophisticated pipeline, using CORN and CORAL to put the different labels in a stricter relation.

Finally, with our proposed methodology using EfficientNet-B5 and CORN we achieve 67.5\% accuracy and a Cohen's Kappa score of 0.451 in predicting the respiratory severity of CMR images. Afterwards, the provided ViT U-Net\textsubscript{2D} is used on samples that do not get classified as having strong respiratory artefacts. ViT U-Net\textsubscript{2D} reaches an average of 86.18\% Dice accuracy on the different labels.

\section{Conclusion}
Detecting respiratory motion in cardiovascular images can help to catch problematic cases where the segmentation model cannot produce high-quality segmentation masks. With the classification model being EffcientNet-B5 + CORN and the segmentation model being ViT U-Net\textsubscript{2D}, we provide a group of tools that allow us to overcome this problem. We catch failure cases of strong severity with the classification model, and the segmentation model allows us to predict high-quality segmentation masks on images with mild to intermediate motion artefacts. We have shown that our proposed workflow can successfully alleviate the daily routine of technicians by providing immediate feedback on image quality during CMR acquisition and therefore saving the hospital time and costs.

\section*{Reproducibility}
Any data split and trained networks will be provided upon acceptance along with instructions on how to run and reproduce all experiments. The code will be made public under \url{https://github.com/MECLabTUDA/QA_med_data/tree/dev_QA_CMRxMotion}.

\section*{Acknowledgements}
This paper has been produced benefiting from the 2232 International Fellowship for Outstanding Researchers Program of TUBITAK (Project No: 118C353). However, the entire responsibility of the publication/paper belongs to the owner of the paper. The financial support received from TUBITAK does not mean that the content of the publication is approved in a scientific sense by TUBITAK.

%
%
%
\bibliographystyle{splncs04}
\bibliography{main.bib}

\end{document}